# Mechanisms of quasi-van der Waals epitaxy of 3D metallic nanoislands on suspended 2D materials

*Kate Reidy[1‡], Joachim Dahl Thomsen[1‡] †, Hae Yeon Lee[1] ††, Vera Zarubin[1], Yang Yu[2], Baoming Wang[1], Thang Pham[1], Priyanka Periwal[3]†††, Frances M. Ross[1*]*

1. Department of Materials Science and Engineering, Massachusetts Institute of Technology, Cambridge, MA 02139, United States.

2. Raith America Inc., International Applications Center, 300 Jordan Road, Troy, NY 12180, USA

3. IBM T. J. Watson Research Center, Yorktown Heights, New York 10598, USA

**Abstract** Understanding structure at the interface between two-dimensional (2D) materials and 3D metals is crucial for designing novel 2D/3D heterostructures and improving the performance of many 2D material devices. Here, we quantify and discuss the 2D/3D interface structure and the 3D morphology in several materials systems. We first deposit facetted Au nanoislands on graphene and transition metal dichalcogenides, using measurements of the equilibrium island shape to determine values for the 2D/Au interface energies and examining the role of surface reconstructions, chemical identity, and defects on the grown structures. We then deposit the technologically relevant metals Ti and Nb under conditions where kinetic rather than thermodynamic factors govern growth. We describe a transition from dendritic to facetted islands as a function of growth temperature and discuss the factors determining island shape in these materials systems. Finally, we show that suspended 2D materials enable the fabrication of a novel type of 3D/2D/3D heterostructure and discuss the growth mechanism. We suggest that emerging nanodevices will utilize such versatile fabrication of 2D/3D heterostructures with well-characterized interfaces and morphologies.

**Main Text**

The integration of 3D metallic nanoislands with 2D materials, to create what we here refer to as 2D/3D heterostructures, enables novel opportunities in plasmonics,[1–3] single photon emission,[4,5] exciton coupling,[6,7] spintronics,[8] and quantum technologies.[9] Moreover, a thorough understanding of the processing and properties of metal deposition on 2D materials is essential for design of devices that incorporate 2D materials. For example, metal nucleation, epitaxy, and defects have a strong influence on 2D/3D device properties such as contact resistance, optical response, and high-frequency electrical performance,[10–19] while metal nanoisland shape also dictates catalytic properties, plasmonic behavior, and excitonic coupling.[1–3,6,7] Shape, nucleation and degree of epitaxy are expected to be controlled by kinetic factors, such as the parameters of the metal deposition on the 2D material, and thermodynamic factors, such as the metal surface energies and the interface energy between the metal and 2D material. As modern combinations of 2D/3D 'mixed dimensional' heterostructures continue to unlock applications in solid-state devices,[1–9] a detailed understanding of how the growth mechanism controls the resulting morphology of the metal and the 2D/3D interface is paramount to their integration into mainstream technologies.[19,20]

The growth of 3D metals on bulk layered materials is well-established and referred to as quasi-vdW epitaxy,[21] to distinguish from vdW epitaxy between 2D materials.[22] The relatively weak quasi-vdW bonding enables sharp interfaces and avoids the requirements of lattice matching in conventional epitaxy.[21] Quasi-vdW epitaxy is well known for facetted nanoislands of Au, Ag, and Au-Pd alloys on graphite;[23–29] Ag, Au, and Cu on molybdenite (bulk $MoS_2$);[30–32] and Au on bulk $WTe_2$, $WS_2$, and $WSe_2$ crystals.[33,34] Scanning tunneling microscopy (STM) studies have probed quasi-vdW epitaxy for numerous metals (Au, Ti, Co, Pd, Pt, Rh, Ir, W, Re, Eu, Gd, Dy, Fe, and Pb)[35–42] on thin but supported 2D materials, such as monolayer graphene on SiC, Ru(0001) and Ir(111), where the substrate underneath the graphene was found to play a defining role in the metal morphology.[12,36,42]

In contrast to these previous studies on bulk and supported 2D materials, epitaxial metal nanoisland growth on thin suspended 2D materials has been challenging due to carbonaceous contamination and gaseous adsorbates.[26,43,44] Suspended (free-standing) 2D materials offer unique opportunities for strain engineering,[45,46] exciton tailoring,[47] nano-electro-mechanical operation,[48] and optical cavities[49,50] which benefit from their integration with metallic nanostructures. Several emerging applications also involve suspended 2D materials to create 3D/2D/3D heterostructures, allowing device geometries such as metal-insulator-metal (MIM) or metal-semiconductor-metal (MSM) junctions.[9,51] Further understanding and control of the suspended 2D/3D and 3D/2D/3D interface therefore holds great promise for solid-state device physics and future integration of 2D and 3D materials.

In this Letter, we experimentally elucidate mechanisms of quasi-vdW epitaxy of single crystalline metal nanoislands with well-defined facetted morphologies on suspended 2D materials. Focusing initially on Au deposited on $MoS_2$ ($MoS_2$/Au heterostructure) as a model system, we investigate epitaxial, facetted nanoisland growth which self-assembles into the equilibrium Winterbottom shape. Cross-sectional imaging shows an Au-S gap of 0.26 nm, in line with expectations for a quasi-vdW $MoS_2$/Au interfacial bond. By measuring the nanoisland shape in three dimensions we provide a quantitative estimate for the 2D/3D interface energy through thermodynamic modelling. We compare these results to Au deposited on graphene, $WS_2$ and $WSe_2$, considering the relationship between the deposited morphology and surface reconstructions, chemical identity, defects and deposition parameters. We then extend our analysis to the technologically relevant metals Ti and Nb, observing a transition from dendritic to compact island shapes with increasing temperature, and we compare the kinetic and thermodynamic factors determining island shape in all these materials systems. Finally, we show that suspended 2D materials enable the formation of metal/2D/metal (3D/2D/3D) heterostructures and consider the growth mechanisms involved. We discuss how an understanding of the structure and energetics of these quasi-vdW interfaces will open opportunities for novel mixed dimensional 2D/3D nanodevices and improve metal integration in current 2D devices.

We first discuss the structure of the $MoS_2$/Au interface, aiming to estimate the quasi-vdW interface energy. Au was deposited on $MoS_2$ under ultra-high vacuum (UHV) conditions (**Methods**) to create islands that are compact triangles ~20 nm in lateral size, flat topped and 4-8 nm in height. **Figure 1a** shows the overall structure and epitaxial alignment, with Au {111} parallel to $MoS_2$ {0001} and Au [220] parallel to $MoS_2$ [$\overline{11}20$], as expected.[15,23,46,52] Cross-sectional high angle annular dark field (HAADF)-STEM imaging suggests that the $MoS_2$/Au interface is atomically abrupt with no detectable chemical mixing (**Figure 1b**), consistent with previous descriptions of weak interactions between metals and $MoS_2$.[18] We measure the Au-S bond length as 0.26 ± 0.02 nm (**Methods**), greater than the sum of the Au and S covalent radii (**Figure 1c**) which indicates the sulfur in MoS$_2$ does not form a strong Au-S bond.[53] Comparison with previous *ab initio* density functional theory (DFT) calculations of relaxed MoS$_2$/Au contacts[53] matches our Au-S distance observation (**Figure 1d,** DFT calculation from Ref. [53]) and predicts a binding energy of ~0.36 eV per atom, on the order of a strong vdW bond.[53] Cross sectional STEM and Raman analysis does not show detectable defect formation in the MoS$_2$ after Au deposition and a maximum of 1% strain within detection limits (**Methods, Figure S1**). Evaporation of Au on MoS$_2$ has been previously shown to result in defective interfaces due to the high kinetic energy of metal atoms.[54] However, defect-free interfaces have been obtained with metals having lower melting points, such as In,[18] or using lower energy deposition methods such as inkjet printing.[55] Therefore, we postulate that the very low rate of our Au deposition (0.05 nm/min) in UHV provides a strategy for minimizing defect formation at the 2D/3D interface.

We use these structural measurements to estimate the quasi-vdW interface energy from thermodynamic modelling of equilibrium island shapes (**Methods**).[56,57] We first confirm that the nanoislands are indeed in the equilibrium shape by observing their stability to high temperature annealing (**Figure 1e,f, Supplementary Movie 1**). On annealing in UHV post-deposition, shapes such as long fingers are not stable (**Figure 1f, black arrow**) however facetted Au triangles remain stable to high temperature (**Figure 1g, red arrow**), confirming they are already at or close to the equilibrium configuration. These nanoisland shapes are consistent with the equilibrium Winterbottom shape of Au on MoS$_2$, which is bounded by {111}, {110} and {100} facets (**Figure 1g,h, Methods,** and **Figures S2-4**) and are consistent across micrometer size areas of the sample (**Figure 1i**).

The equilibrium shape is determined by the interfacial energies between the particle and its environment ($\gamma_{PV}$, where P refers to the particle and V refers to the environment, here vacuum), the 2D material and environment ($\gamma_{SV}$) and the particle and 2D material ($\gamma_{SP}$).[56,57] Specifically, the height of the Winterbottom shape is determined by the parameter ($\gamma_{SP}$ - $\gamma_{SV}$)/ $\gamma_{PV}$. We used AFM to obtain upper and lower bounds for the heights of several islands (**Figure S3**), confirmed more precisely by cross-sectional STEM (**Figure 1g**). The outcome (see derivation in **Methods**) is an estimate of the energy ratio,

$$\frac{\gamma_{SP} - \gamma_{SV}}{\gamma_{PV}} \approx -0.14 \pm 0.03 \quad (1)$$

The appropriate $\gamma_{PV}$ is the Au {111} surface energy (0.71 J/m²) while $\gamma_{SV}$, the surface energy of the (0001) MoS$_2$ surface, is obtained from literature as ~0.24 – 0.28 J/m².[58–60] Inserting these values for $\gamma_{PV}$ and $\gamma_{SV}$ into Equation (1) predicts an interface energy for MoS$_2$/Au ($\gamma_{SP}$) of 0.14 – 0.18 J/m². The uncertainty in this estimate depends on factors such as the exact facet configuration, environment, interfacial reconstructions, and temperature.[61] However, we note that the estimated value obtained is close in magnitude to half the interfacial adhesion energies of MoS$_2$/MoS$_2$ (0.482 J/m²)[62] and MoS$_2$/SiO$_2$ (0.17 – 0.42 J/m²)[62] suggesting, as may be expected,[34] that the energy of this 2D/3D quasi-vdW interface is similar to that of the interface between vdW layers.

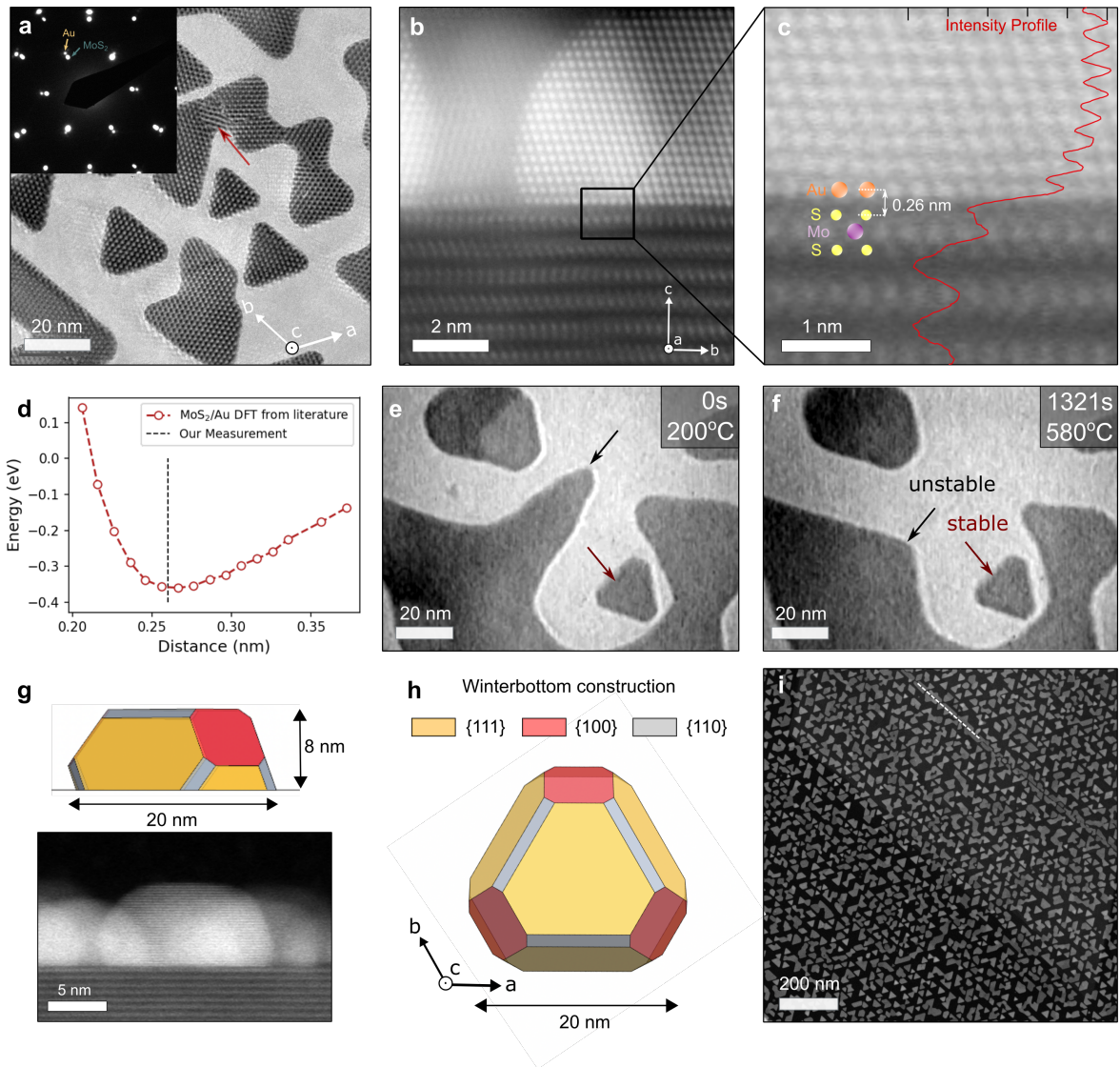

**Figure 1: Quasi-vdW epitaxial interface of facetted Au nanoislands on MoS$_2$. a,** Au nanoislands after deposition on suspended 5L MoS$_2$ that had been annealed in ultra-high vacuum at ~500 °C for several hours prior to deposition. One epitaxially misaligned region is observed in the image, which we attribute to island coalescence (red arrow). Inset shows an epitaxially aligned crystalline diffraction pattern. **b,** Cross sectional STEM image of the MoS$_2$/Au interface. Slight distortion of MoS$_2$ lattice is thought to be due to lattice disturbance during TEM cross section preparation. **c,** Atomic resolution HAADF-STEM image of the interface with overlaid averaged vertical intensity profile and schematic. Au-S distance calculated statistically from intensity peaks

as 0.26 ± 0.02 nm (**Methods**). Mo-Mo and Au-Au in plane distances are calculated from the same image as 0.316 ± 0.02 nm and 0.291 ± 0.01nm, respectively. **d,** Comparison of our experimentally determined Au-S bond distance with theoretically calculated relaxed contact regions between MoS$_2$/Au(111). DFT calculation data from ref [53] **e,f,** Stills from movie showing *in situ* annealing of MoS$_2$/Au. The time and temperature of each image is indicated. Zero time corresponds to the start of heating, which was achieved by passing a current through the chip to heat it resistively, with temperature read using a calibration curve measured in vacuum using an infrared pyrometer. **g,** Side view of Winterbottom construction with $(\gamma_{SP} - \gamma_{SV})/\gamma_{SP}$ = -0.14 (top) and corresponding cross-sectional STEM image showing MoS$_2$/Au island shape (bottom). **h,** Theoretical plan-view Winterbottom construction showing the equilibrium shape of Au nanoislands on MoS$_2$. **i,** Overview image of facetted nanoparticles on suspended MoS$_2$ over micrometer area. White dashed line indicates location of a step edge on the MoS$_2$ decorated with Au. Islands are observed to preferentially decorate step edges and folds on the 2D material surface; such preferential nucleation has been attributed to the low binding energy coupled with fast diffusion of metal atoms on 2D material terraces.[14,24] Note the nucleation density and island shape is the same on either side of the step edge, indicating that the layer number of MoS$_2$ has negligible influence on Au growth kinetics.

We now compare these structural and energetic measurements for the MoS$_2$/Au quasi-vdW interface to the results of depositing Au nanoislands on other 2D materials: Gr (**Figure 2a-c**), WSe$_2$ (**Figure 2d**), and WS$_2$ (**Figure 2e**). Au exhibits quasi-vdW epitaxy on all the 2D materials examined, with majority single crystalline islands except for occasional stacking faults from coalescence boundaries (**Figure 2c, inset**).

On Gr, the Au exhibits a more pointed equilibrium shape than on $MoS_2$ (see the triangle corners in **Figure 2a vs. Figures 1a and 2d,e**). We suggest that this is due to reconstruction of the Au(111) surface. Au(111) is well known to form a herringbone surface reconstruction in UHV,[63] which can decrease the energy of the Au(111) facet by up to ~1.95 eV/unit cell.[64] Image contrast consistent with the herringbone structure is readily observed using dark field TEM imaging (**Figure 2b**) and integrated differential phase contrast (iDPC) STEM imaging (**Figure 2c**). Incorporating this lower surface energy results in an equilibrium Winterbottom shape with a higher percentage of Au(111) top surface plane and hence more pointed tips (**Figure 2a inset, Methods**). We model $\gamma_{PV}$ as an estimated Au{111} reconstructed surface energy (0.56 J/m$^2$) and obtain the surface energy of Gr ($\gamma_{SV}$) from literature (~0.19 – 0.26 J/m$^2$).[65] Incorporating these values with the measured island shape yields a calculated Gr/Au interface energy ($\gamma_{SP}$) of ~ 0.07 – 0.14 J/m$^2$ (**Methods**), slightly lower than that obtained for the $MoS_2$/Au system. This may be expected due to the difference in strength between Au-C and Au-S bonding.[66,67]

On $WSe_2$ and $WS_2$, the Au exhibits a crystal shape similar to that on $MoS_2$. This is corroborated by TEM imaging at 20° tilt angle to show the crystal facets (**Figure 2d,e, insets**). Furthermore (**Figure 2d,e**), we do not observe contrast suggesting surface reconstructions in Au islands on any of the TMDs ($MoS_2$, $WSe_2$ and $WS_2$). These results are consistent with an unreconstructed Au(111) surface in the TMD/Au system, and indeed the herringbone reconstruction is not expected to be present upon interaction with sulfur.[68] Furthermore, we also note that the average island size on TMDs is smaller than on Gr (note the same scale bar in **Figures 2a vs. 2d,e**), which decreases the energy benefit and hence probability of forming surface reconstructions.[63]

Statistical analysis (**Methods**) shows that Au on Gr and Au on TMDs differ not just in island shape, but also in nucleation density, island size and quality of the epitaxy. We find higher disorder, lower nucleation density, and larger average size for Au islands on Gr (**Figure 2f**) compared to Au on the TMDs, when deposited under the same conditions of total flux, flux rate and temperature. We have discussed above how the equilibrium shape depends on the interface energy and therefore, ultimately, the different strengths of the Au-S, Au-Se and Au-C interactions. We also expect interfacial interactions to influence ordering, nucleation density, and island size. For example, weak interfacial interactions obviously reduce the energy penalty for poor alignment and hence the quality of epitaxy. Other factors also connect the interface character with the shape, nucleation density, size and degree of disorder. Defect density and defect chemical properties are well known to influence metal nanoisland nucleation and morphology.[32,34] Coalescence defects such as grain boundaries and dislocations are more likely in islands that have nucleated close together, and islands with such defects may show characteristic non-Winterbottom shapes,[67] as in the example in **Figure 2c**. Finally, nucleation density may influence shape since higher nucleation density is associated with smaller islands (for constant total deposited amount) and islands that are smaller may lack surface reconstructions, as discussed above.

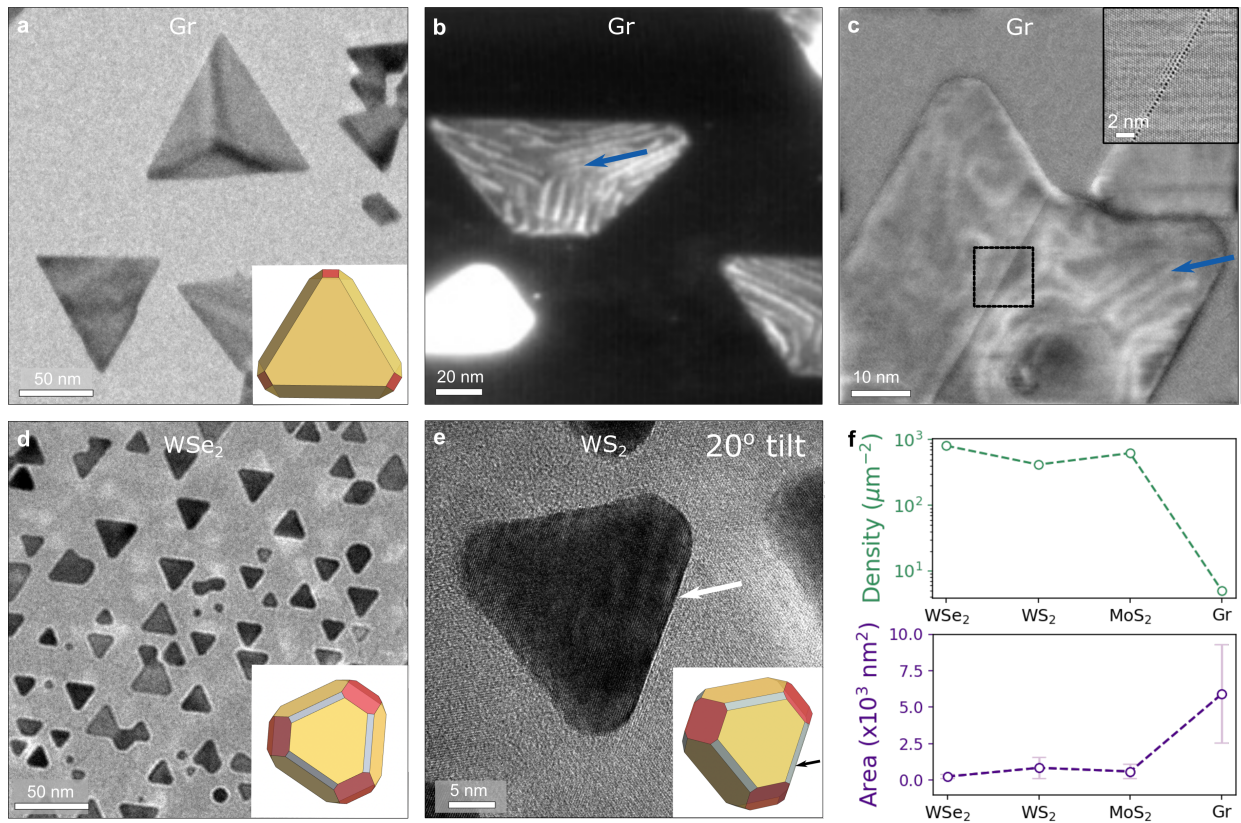

**Figure 2: Comparisons of Au nanoisland growth on several 2D materials. a,** TEM image of Au nanoislands on clean suspended Gr. Inset shows theoretical Winterbottom construction with Au(111) surface energy reduction by 0.15 J/m$^2$ to account for surface reconstruction. The islands are curved out of plane (shaped like shallow bowls) as seen from bend contours in TEM, especially when larger. **b,** Dark field TEM image of Au(111) surface reconstruction in UHV. The bright curved features (blue arrow) are caused by the strain fields of the herringbone surface reconstruction of Au(111). **c,** Integrated differential phase contrast (iDPC) STEM image of defective Au nanoisland on Gr. The bright curved features (blue arrow) are caused by the strain fields of the herringbone surface reconstruction of Au(111). Inset shows iDPC STEM image of stacking fault caused by island coalescence. Such coalescence defects can result in island shapes that differ from the equilibrium morphology.[69] **d,** TEM image of Au nanoislands on TMD WSe$_2$. Inset shows Winterbottom construction. **e,** 20° tilt TEM image of Au nanoisland on TMD WS$_2$,

inset shows corresponding Winterbottom shape at 20 ° tilt. White arrow indicates sharp side facet. **f,** Statistical analysis of nucleation density (top) and island area (bottom) of Au deposition on Gr and TMDs. Error bars in area represent the standard deviation. Nucleation density data points are measurements from a single suspended area.

Since the nature of the initial 2D surface plays a key role in determining both interface energy and nucleation density, we next explore the degree of control that is possible by carrying out Au deposition on surfaces that have different types of defects, controlled either by deliberate patterning or through tailoring the 2D preparation conditions.

**Figure 3a-d** illustrates the deposition of Au on Gr in which irradiation with $Si^{++}$ ions was used to create defects. Nucleation density increases and island size decreases with dose. This is consistent with a mechanism of preferential nucleation at the ion-induced defects. The island morphology also depends on defect density. Small polycrystalline islands form that are not in their facetted equilibrium shape, indicating a loss of epitaxial alignment and a lower diffusion distance. This can also occur with other heterogeneities, such as carbon contamination and step edges, as shown in **Figure S5**. In a second series of depositions, this time on $WS_2$, we found that the Au nucleation density and morphology were sensitive to the density of defects on the surface. The preparation of the $WS_2$, described in **Figures S6-S7** and discussed further in the **Supplementary Information,** yielded a lateral concentration gradient of defects that could be measured by photoluminescence. All of these observations suggest that engineering the defect density and chemical/physical heterogeneities across a 2D flake may allow control of nucleation density, size, and faceting of islands grown on the surface.

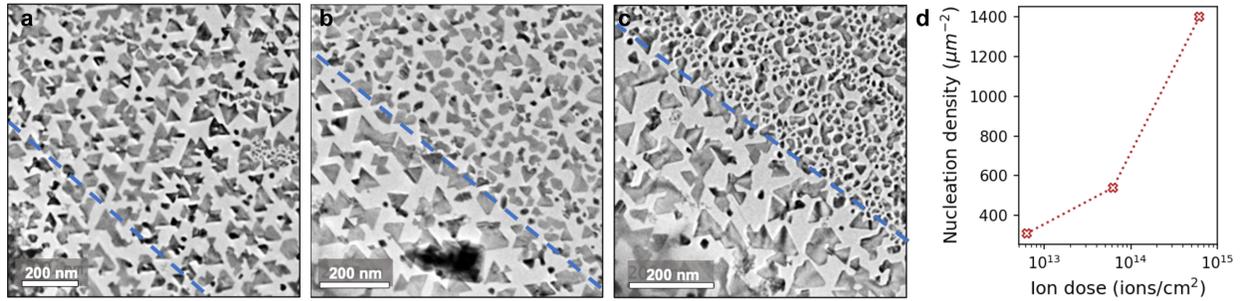

**Figure 3: Role of defects in island nucleation. (a-c)** Au nanoisland growth on Gr that had previously been irradiated by $Si^{++}$ ions with an array of point shots with X-Y spacing of 50 and 100 nm with **a,** 6.24 x $10^{12}$ ions/cm$^2$, **b,** 6.24 x $10^{13}$ ions/cm$^2$, and **c,** 6.24 x $10^{14}$ ions/cm$^2$, respectively. Areas above the blue dashed line are the exposed areas, and below are unexposed. **d,** Plot of nucleation density vs. ion dose for the irradiated areas shown in a, b, c.

We now discuss 2D/3D growth for two other metals important in electronic contacts and superconducting circuits, Ti and Nb respectively (**Figure 4**). Ti and Nb are deposited via electron beam evaporation in UHV (**Methods**). The results we obtain for these metals, strikingly different from Au, can be understood through the influence of kinetic factors rather than the thermodynamic control we deduced for the Au island growth. The Ti and Nb morphologies show a signature of a kinetically limited growth mechanism (fractal-dendritic growth[70,71]). Moreover, the growth morphology varies distinctively with the 2D material temperature. In **Figure 4e-l**, Ti and Nb exhibit a morphology transition on raising the growth temperature that is consistent with literature of temperature-dependent and annealed metal growth on bulk graphite.[25,29,72,73] The temperature required for equilibrium nanoisland geometries is expected to depend on the vapor pressure and melting point ($T_m$) of the metal. We obtain facetted shapes at relatively high temperatures for Ti (~550 °C) and Nb (~900 °C). Even for Au, where we obtained the equilibrium Winterbottom shapes at only 50-100 °C (**Methods**), temperatures lower than this showed hints of dendritic morphology (**Figure 4a-d**). From these three sets of experiments it appears that 2D/3D fractal-dendritic growth occurs on the order of ~0.1 $T_m$, whereas > 0.3 $T_m$ is required for facetted growth, in agreement with requirements for homoepitaxy of metallic islands.[71] We are unable to reach the high temperatures we anticipate would be required to create Ti and Nb islands in their full Winterbottom shape (**Figure S4)** and our Ti and Nb island shapes remain consistent with kinetically limited growth (**Figure 4l**).

It is interesting to note that strong epitaxy is visible even for dendritic Ti and Nb structures deposited on suspended layers at room temperature, as shown in diffraction (**Figure 4e,i, insets)**. Therefore, the temperature required for epitaxy is lower than that required for facetted islands. Although the epitaxial alignment at room temperature is not as strong as at higher temperature (**Figure 4e** vs. **f** or **Figure 4i** vs. **j insets**), for the purposes of creating epitaxially aligned films a thick room temperature deposition should suffice if the 2D material is prepared as described above. We also find this to be true of metals deposited on supported 2D materials, where facetted island shapes are not observed but epitaxy can be maintained (**Figure S8**). This result is relevant in forming continuous films without large angle grain boundaries, and is in agreement with literature of epitaxial thin film growth on layered materials at room temperature.[74–76]

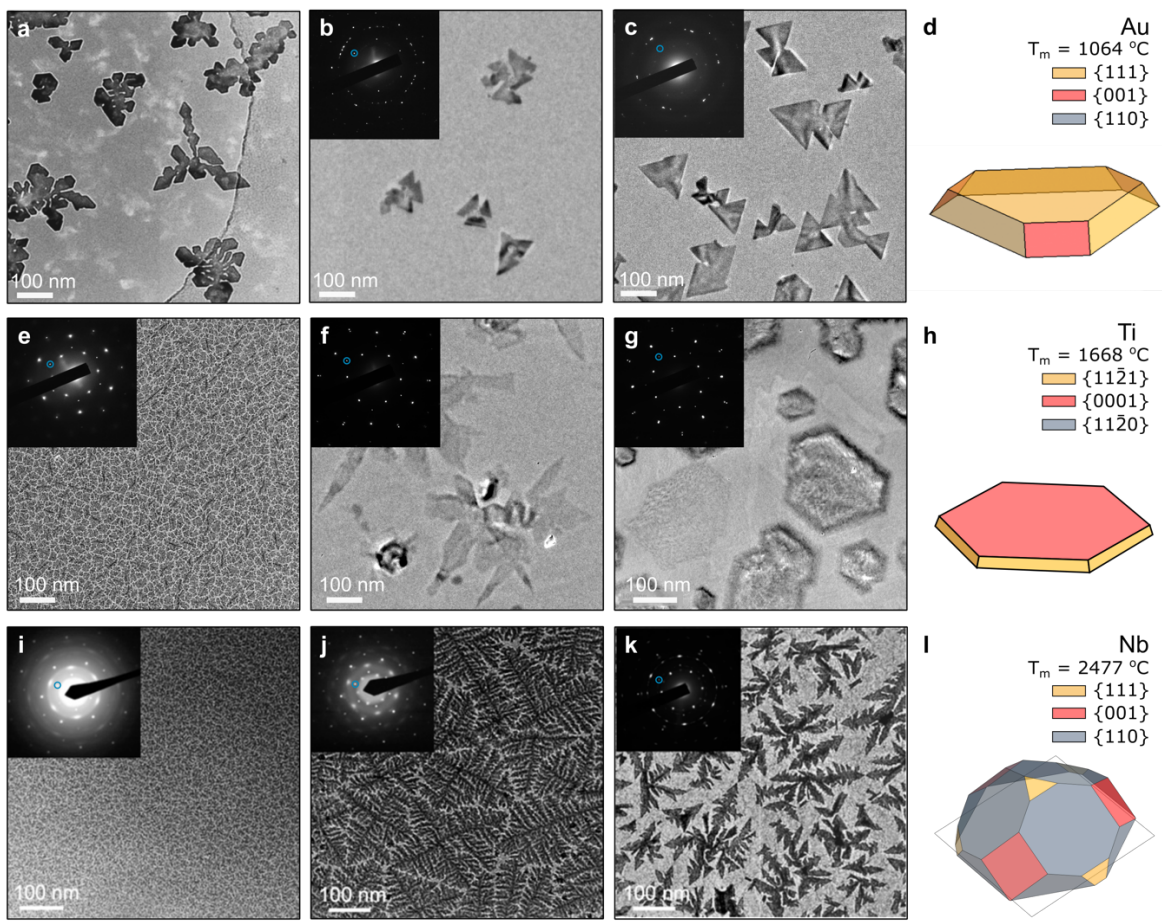

**Figure 4: Temperature effects and facetted growth of Au, Ti, and Nb nanoislands on Gr.** (**a-c**) Au growth on suspended few-layer Gr. No intentional heating was used, but different sample temperatures were obtained by varying the distance between the sample and the source, and thermocouple measurements suggest the sample temperature is 50-100 ºC during evaporation. **a,** dendritic morphology (deposited at a lower temperature than **b** and **c** with the sample further from the source) **b,** compact dendrites, and **c,** equilibrium shapes (both estimated at 50-100 ºC). **d,** Calculated equilibrium Winterbottom shape of Au on Gr. (**e-g**) Ti growth on suspended few-layer Gr at **e,** room temperature, **f,** ~300 ºC, and **g,** ~550 ºC. **h,** Calculated equilibrium Winterbottom shape of Ti on Gr. (**i-k**) Nb growth on suspended few-layer Gr at **i,** room temperature, **j,** ~600 ºC, and **k,** ~900 ºC. Island shape remains dendritic and far from the calculated shape. **l,** Calculated

equilibrium Winterbottom shape of Nb on Gr. Insets in TEM images show corresponding epitaxial diffraction patterns, where the Gr($10\bar{1}0$) spots are circled in blue.

We finally discuss an extension of the deposition strategies we have shown above. Some applications of 2D/3D integration benefit from the ability to deposit metal nanoislands on both surfaces of a suspended 2D material. The 3D/2D/3D structures suggest opportunities for device geometries such as capacitors, metal-insulator-metal tunnel junctions, and Josephson junctions employing 2D materials as the tunnel barrier (e.g. Al/MoS$_2$/Al or Nb/MoS$_2$/Nb).[9,51] **Figure 5a-c** demonstrates such a 3D/2D/3D geometry, where the top and bottom islands are epitaxially aligned with the 2D material (**Figure 5b**). Some island pairs show small rotations (e.g. of 1-2º) which result in the formation of moiré patterns (**Figure 5c**). This rotation is within the statistical variation of Au island rotation on one-sided Gr deposition (0-4º, measured statistically), and is higher than the rotation variation of Au on MoS$_2$ (~0.3º, determined by moiré periodicity[15]) consistent with the weaker interaction of Au with C than with S/Se. We find that these top/bottom alignments are statistically more likely to occur on monolayer 2D material (**Figure 5d**) than on bilayer (**Figure 5e**) or multilayer (**Figure 5f,g**). We attribute the 3D/2D/3D monolayer correlation to either defect sites that nucleate a pair of top and bottom islands, or nucleation in an inhomogeneous strain field caused by the first-grown set of islands deforming the Gr. We do not expect remote epitaxy through vdW substrates to be responsible for the alignment since polarity is required for this mechanism to operate.[77] The pristine and epitaxial 3D/2D/3D interfaces created in UHV are predicted to improve electrical contact resistance,[11] and provide a platform to study further electrical properties, such as tunnelling current and conductivity, as a function of moiré potential in these 3D/2D/3D heterostructures.[15]

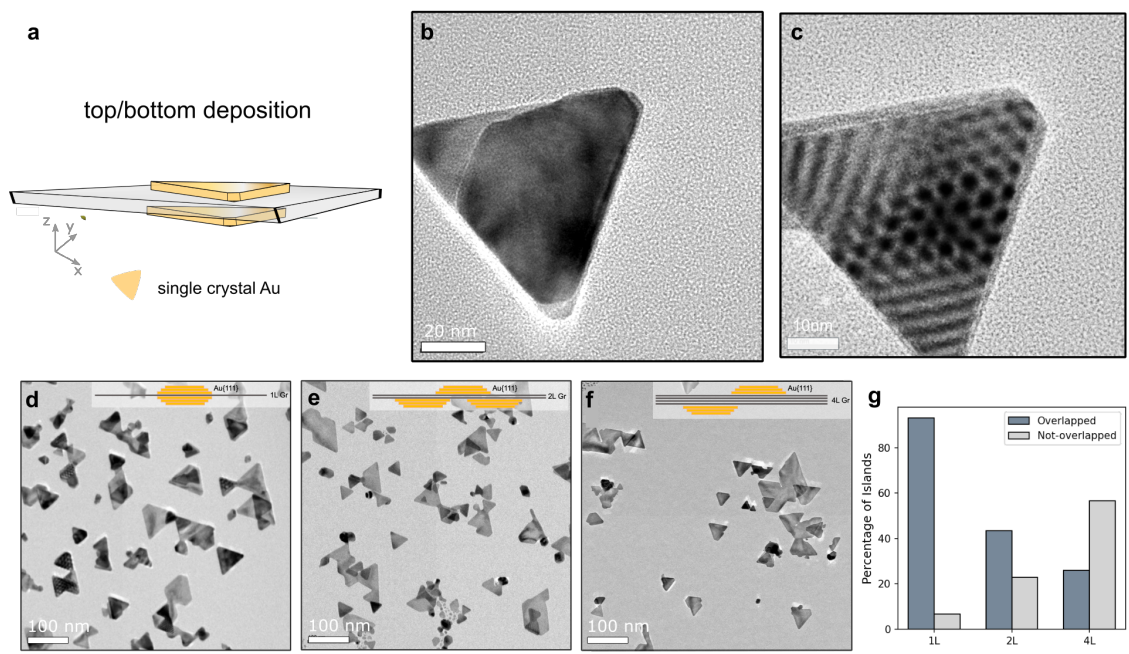

**Figure 5: Top/bottom deposition on suspended 2D materials. a,** Schematic of single crystalline Au nanoisland growth on both the top and bottom of monolayer 2D material. **b,** TEM image of Au nanoisland growth on both sides of monolayer graphene that had been annealed in ultra-high vacuum at ~500 ºC for several hours prior to deposition. Overlapped islands are perfectly epitaxially aligned. **c,** TEM image of Au nanoisland growth on both sides of monolayer graphene that had been annealed in ultra-high vacuum at ~500 ºC for several hours prior to deposition. Moiré patterns visible from two overlapping Au islands show a rotation of 1-2 degrees. **d-f,** TEM images of Au top/bottom deposition shown in **d,** for monolayer, **e,** for bilayer, and **f,** for four layers. Insets show schematic of deposition. **g,** Corresponding statistical analysis of percentage of overlapped vs. non-overlapped islands for monolayer, bilayer, and four layer graphene.

In summary, we have examined a variety of 2D material/3D metal systems (Au on $MoS_2$, Gr, $WSe_2$, and $WS_2$, and Ti and Nb on Gr) to elucidate the relationships between 2D/3D interface structure and energy and nanoisland nucleation and morphology, including the importance and opportunities arising from defect engineering. For the $MoS_2$/Au model system, we quantitatively estimated the quasi-vdW interface energy by correlating microscopy with thermodynamic modelling and we demonstrated that low deposition rates in UHV produce interfaces without detectable defects or intermixing. By comparing different materials systems and growth conditions we delineated conditions where thermodynamic control of island shapes is possible. We also showed that suspended 2D materials enable the fabrication of a novel type of 3D/2D/3D heterostructure.

We predict the thermodynamic and kinetic relationships and measurement strategies described here may be applicable to metals that have been previously found to exhibit epitaxy on bulk or supported vdW materials: Cu, Ag, Pd, Co, Rh, Eu, Gd, Dy, and Pb, among others.[30–32,35–38,40–42] *Ab initio* calculations also predict that most of the *3d* and group 10 transition metals, noble metals, and rare earth metals should exhibit a 3D growth mode on Gr,[78] a result our studies suggest could be more broadly applied to other 2D materials. This opens boundless combinations of metals and 2D materials available for facetted 2D/3D heterointegration.

We finally note that a benefit of the epitaxial suspended 2D/3D and 3D/2D/3D platforms we have examined here is the ability to perform correlative structure and property measurements on the same sample. For example, S/TEM is well suited to visualize the interface and provide atomic-scale strain mapping.[63] Nanoscale angle resolved photoelectron spectroscopy (nano-ARPES),[79] performed from the back surface, could probe the effect of metal deposition on 2D material monolayers. Standard electronic property measurements could also be performed by post-transfer of suspended devices onto electrically contacted substrates, or fabrication of suspended heterostructures with graphene acting as gate electrodes. We hope this work provides general insight to coupling 3D nanoislands with suspended 2D materials and aids in the efficient integration of 2D/3D and 3D/2D/3D heterostructures in future electronic, magnetic, and optical nanodevices.

ASSOCIATED CONTENT

**Supporting Information**. The following files are available free of charge in the Supplementary Information (PDF). Methods: including preparation of the suspended 2D material, ion-beam irradiation, photoluminescence characterization, UHV deposition, TEM and STEM imaging, AFM, Winterbottom construction, Raman spectroscopy, and statistical image analysis; Supplementary Movie 1: *In situ* heating of Au on $MoS_2$; Supplementary Figures 1-8: describing Raman spectroscopy, equilibrium shapes, role of heterogeneities in island deposition, atomic resolution STEM imaging, variation of parameter space, and supported deposition.

AUTHOR INFORMATION

**Corresponding Author**

*fmross@mit.edu

**Present Addresses**

†John A. Paulson School of Engineering and Applied Sciences, Harvard University, Cambridge, Massachusetts 02138, USA.

†† Department of Mechanical Engineering, Columbia University, New York, NY, USA

††† Bureau of Economic Geology, The University of Texas at Austin, University Station, Box X, Austin, Texas 78713-8924, USA

**Author Contributions**

K.R., J.D.T., and F.M.R conceived the project. K.R. and J.D.T. developed the epitaxial deposition, fabricated samples, and performed TEM imaging. K.R. performed atomic resolution and cross-sectional STEM imaging, thermodynamic modelling of Winterbottom constructions, *in situ* heating experiments, and statistical image analysis. J.D.T. performed AFM and Raman measurements. H.Y.L. grew CVD $WS_2$ and performed photoluminescence measurements. K.R., V.Z., and Y.Y. performed VELION FIB defect creation. P.P. performed dark field imaging. T.P. performed iDPC imaging. B.W. prepared FIB cross sections. K.R. wrote the manuscript, with input from J.D.T., F.M.R and all authors. All authors have given approval to the final version of the manuscript. ‡These authors contributed equally.

ACKNOWLEDGMENT

This work was carried out with the use of facilities and instrumentation supported by NSF through the Massachusetts Institute of Technology Materials Research Science and Engineering Center DMR - 1419807. This work was carried out in part through the use of MIT.nano's facilities. This work was performed in part on the Raith VELION FIB-SEM in the MIT.nano Characterization Facilities (Award: DMR-2117609). K.R. acknowledges funding and support from a MIT MathWorks Engineering Fellowship and ExxonMobil Research and Engineering Company through the MIT Energy Initiative. J.D.T. acknowledges support from Independent Research Fund Denmark though Grant Number 9035-00006B. V. Z. acknowledges funding from the Microscopy Society of America Undergraduate Research Scholarship. The authors would like to acknowledge Michael Tarkanian for help in manufacturing TEM sample holders; Dr. Aubrey Penn and Russ Henry for advice on STEM imaging; Prof. Rafael Jaramillo, Prof. Craig Carter, Prof. Carl Thompson, Dr. Max L'Etoile, and Dr. Georgios Varnavides for useful discussions on Winterbottom shapes and interface energies; Dr. Rami Dana for help with UHV equipment; Dr. Julian Klein and Sara Sand for input on nucleation density; and Prof. Jeehwan Kim for equipment access. Reprinted modified Figure 1d with permission from: Igor Popov, Gotthard Seifert, and David Tománek, Physical Review Letters, 108, 156802, 2012. Copyright 2012 by the American Physical Society.

**TOC Graphic**

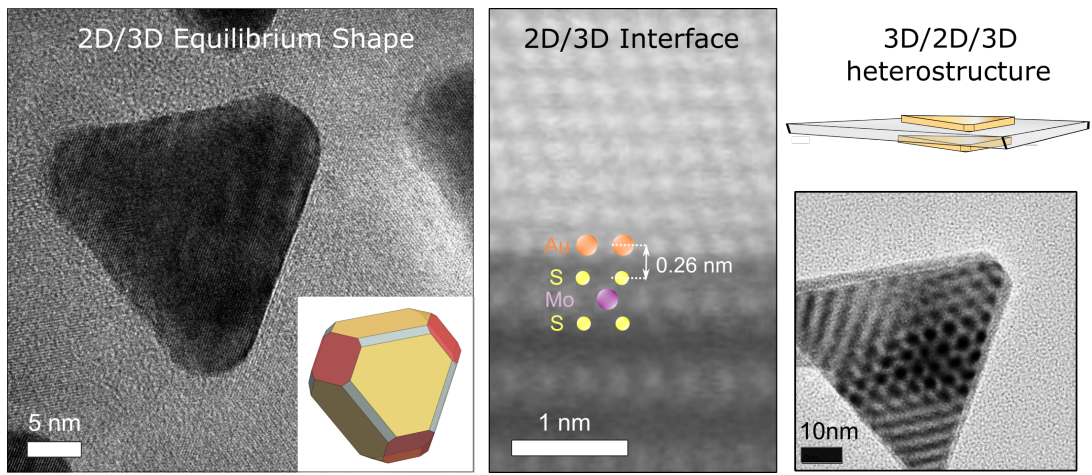